\documentclass[aps,pra,twocolumn,superscriptaddress,showpacs]{revtex4}
\usepackage{amsmath,amssymb,graphicx,color}
\usepackage{bm}
\usepackage{mathrsfs}

\begin{document}
%\draft
%\twocolumn[\hsize\textwidth\columnwidth\hsize\csname@twocolumnfalse\endcsname
\title{Approaching Infinite Temperature upon Repeated Measurements of a quantum system}

\author{Juyeon Yi}
\affiliation{Department of Physics, Pusan National University, Busan 609-735, Korea}
\affiliation{Institut f\"{u}r Physik, Universit\"{a}t Augsburg, Universit\"{a}tsstra\ss e 1, D-86135 Augsburg, Germany}
\author{Peter Talkner}
\author{Gert-Ludwig Ingold}
\affiliation{Institut f\"{u}r Physik, Universit\"{a}t Augsburg, Universit\"{a}tsstra\ss e 1, D-86135 Augsburg, Germany}

\date{\today}
\begin{abstract}
The influence of repeated projective measurements on the dynamics of the state
of a quantum system is studied as a function of the time lag $\tau$ between successive measurements. In the limit of infinitely many measurements of the occupancy of a single state the total system approaches a uniform state.
The asymptotic approach to this state is exponential in the case of finite Hilbert space dimension. The rate characterizing this approach
undergoes a sharp transition from a monotonically increasing to an erratically varying function of the time between subsequent measurements.
\end{abstract}
\maketitle

\section{introduction}
Even though the theory of measurements provides an integral part of Quantum Theory and even though to this day it has passed any close scrutiny the measurement process remains a topic of controversial discussions. According to von Neumann \cite{vN} the measurement of an observable yields one of the eigenvalues of the hermetian operator representing the measured observable and projects the state of the system onto the eigenspace of the resulting eigenvalue. There exist several attempts to explain this state reduction as resulting from the interaction of the system with a measurement apparatus whereby the system combined with the environment undergoes a unitary time evolution. Depending on the special structure of the measurement apparatus and its coupling to the system the reduced system state undergoes decoherence~\cite{dephasing1, dephasing2,dephasing3,dephasing4} with respect to the eigen-basis of the operator to be measured. At the same time the state of the measurement apparatus adopts a particular ``pointer state''. The latter process, also known as ``ein\-se\-lection'', makes it possible to read out the result of the measurement~\cite{otherapproach1, otherapproach2,otherapproach3}.

As for other physical processes, any measurement resulting from an interaction of the system with the measurement apparatus will take a finite amount of time. This aspect is not taken into account in von Neumann's idealized picture where a projective measurement is considered as an instantaneous process. It is therefore conceivable to consider arbitrarily frequent measurements which will lead to a total freezing of the dynamics, a phenomenon that is known as the Zeno effect~\cite{zeno1, zeno2, zeno3, zeno4, KS, FP}. When a finite, but still sufficiently short time $\tau$ elapses between two measurements of the same observable the probability to consecutively find the system in the same eigenstate of the observable exponentially decreases with a rate that is proportional to the time $\tau$ separating two measurements. Experimental verifications of the Zeno effect are reported in Refs.~\cite{smexptwolevel, Kwiat}.
Most recently, direct consequences of projective measurements have been discussed in different contexts such as thermodynamic control~\cite{erez}, quantum tomography~\cite{navascues}, and fluctuation theorems~\cite{campisi}. 

In this work we investigate the combined action of unobserved quantum motion and repeated projective measurements.
In contrast to the standard treatment of the Zeno effect, according to which the system is followed only until it has left the initial state for the first time
we keep track of all possible outcomes up to a total observation time. In this way we take into account also realizations with intermediate deviations from and possibly multiple revivals of the initial state. This approach differs from the standard treatment of the Zeno effect imposing selective measurements \cite{vN}, which disregards the future fate of the system after it has left the initial state for the first time.
We find for systems with finite-dimensional Hilbert space that asymptotically, in the limit of infinitely many measurements, all states of the system acquire equal probability. Hence, the probability to retrieve the initial state is given by the inverse of the Hilbert space dimension. This result is universal and in particular does not depend on the time $\tau$ between two consecutive measurements as long as $\tau$ is different from zero. Exceptions from this universal behavior are found only for those singular values of the time between measurements and of the system parameters for which an invariant state exists of that part of the system that is not directly effected by the measurement.
The approach to the stationary state eventually becomes exponential.
The corresponding decay rate is a monotonically increasing function of $\tau$ up to the time scale $\tau^*$ corresponding to the inverse of the largest energy difference of the system.
For times larger than $\tau^*$ the decay rate shows a highly irregular behavior.

This paper is organized as follows: Sec.~II is devoted to the
introduction of the system and the measurement of interest. In Sec.~III, we investigate the time evolution of the density matrix under the repeated measurements and point out the universality of the stationary state of equal probability, followed by a more detailed discussion and proof in Sec.~IV. We then focus on the survival probability
in Sec.~V, and its decay behavior is detailed in Sec.~VI. The paper ends with a summary and conclusions %are drawn
in Sec.~VII. 

%%%%%%%%%%%%%%%%%%%%%%%%%%%%%%%%%%%%%%%%%%%%%%%%%%%%%%%%%%%%%%%%%%%%%%%%%%%%%%%%%%%%%%%%%%%%%%%
\begin{figure}[t]
\includegraphics[width=.8\columnwidth]{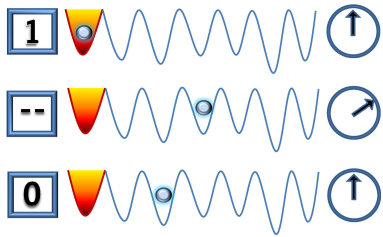}
 \caption{(Color online)~Illustration of the system and the measurement protocol: The counter displays `1' when the particle occupies the left-most colored site, referred to as dot. When the particle
resides at any chain site represented by one of the wells, the counter reports the event as `0'.
Each time the one-handed clock has completed a period $\tau$ a projective measurement is performed and its result displayed by the counter which remains idle in the meanwhile.
 }\label{sche}
\end{figure}
%%%%%%%%%%%%%%%%%%%%%%%%%%%%%%%%%%%%%%%%%%%%%%%%%%%%%%%%%%%%%%%%%%%%%%%%
\section{The System}

We consider the dynamics of a single particle on a lattice consisting of a single site,
referred to as dot and % coupled to a reservoir made of
a one-dimensional chain with $N$ sites acting as a reservoir.
A schematic view of this system at different stages during a measurement cycle 
is presented in Fig.~1. Any state of the particle can be represented in terms of linear combinations of the discrete position states $|x\rangle $ where $x = 0$ indicated by the coloring in Fig.~\ref{sche} denotes the dot position and $x=\ell$ with $\ell=1\ldots N$ the chain positions, respectively. The total system is described by the Hamiltonian
\begin{equation}
\begin{split}
H &= \sum_{\ell=1}^N \epsilon_\ell  |\ell \rangle \langle \ell |
 - \gamma_{c}\sum_{\ell=1}^{N-1}\left (|\ell\rangle \langle \ell+1|
+|\ell+1\rangle \langle \ell|\right )\\
&\quad - \gamma \left ( |0\rangle \langle 1|+|1\rangle \langle 0| \right )\:,
\end{split}
\label{H}
\end{equation}
where $\epsilon_\ell$ denotes the on-site energy of the site $\ell$ relative to the dot-site energy $\epsilon_0=0$. Transitions between neighboring sites of the chain are governed by the parameter $\gamma_{c}$ which will be used as energy unit, hence $\gamma_{c}=1$.
Here we assume that only one end~$(\ell=1)$ of the chain is
coupled to the dot. This model Hamiltonian has various
applications. On one hand, it can be interpreted in terms
of the Wigner-Weisskopf model~\cite{ww} for the decay of the
excited state of a two-level atom coupled to a finite number of electromagnetic modes
of an empty cavity. On the other hand, it describes a quantum dot coupled to a one-dimensional
wire and, moreover, the dynamics of a cold atom in a one-dimensional optical lattice. Recently the decay of an unstable quantum state  was discussed in terms of 
the Hamiltonian
%to Eq.
~(\ref{H})
with vanishing on-site energies $\epsilon_\ell =0$, identical hopping rates $\gamma_c =\gamma$ and infinite $N$~\cite{longhi}. In contrast to our present study effects of repeated measurements were not taken into account in this study.

We consider the measurement of the occupancy of the dot-state that is represented by the
projection operators onto the dot~($d$) and the chain~($c$), respectively,
\begin{equation}
\Pi_{d}=|0\rangle \langle 0|\:, \hspace{0.2cm}
\Pi_{c}=\sum_{\ell=1}^N|\ell\rangle \langle \ell| \:.
\end{equation}
Note the completeness,
$\Pi_{d}+\Pi_{c}=1$, and the orthogonality, $\Pi_i^\dagger = \Pi_i$, $i=c,d$,  $\Pi_{d}\Pi_{c}=0$ of the projection operators. For chains with more than a single state, $N>1$, the observation of the dot-state occupancy corresponds to a selective measurement in the sense that the state reduction only removes dot-chain correlations but leaves the internal chain part of the state unchanged.

\section{Density Matrix}

 We denote the density matrix of the total system
immediately after the $M$-th measurement by $\rho_{M}$. The discrete time dynamics transforming
the density matrix from the moment immediately after the $M$-th measurement to the moment immediately after the $M+1$-st measurement,
\begin{equation}\label{it}
\rho_{M+1} = \mathcal{M} \rho_M\:,
\end{equation}
is governed by the linear map $\mathcal{M}$ given by
\begin{equation}
\mathcal{M} \rho =\sum_{\alpha=c,d}\Pi_{\alpha}U(\tau)\rho U^{\dagger}(\tau)\Pi_{\alpha}\:.
\label{M}
\end{equation}
It is composed of the unitary time evolution generated by
$U(\tau) = \exp \left (-i H \tau/\hbar \right )$
and a subsequent measurement-induced state reduction onto the subspaces of dot and chain. This reduction cancels any quantum correlation between dot and chain at each measurement by setting the nondiagonal matrix elements $\langle 0|\rho|\ell \rangle$ with $\ell =1 \ldots N$ to zero.
Hence, all solutions of (\ref{it}) are of the form
\begin{equation}\label{az1}
\rho_{M}=p_{M}\Pi_{d}+\chi_{M}\Pi_{c}\:,
\end{equation}
where $p_{M}$ is a real number, while $\chi_{M}$ is a reduced density matrix of the  chain.
Equation (\ref{it}) then decomposes into the following recursion relations for $p$ and $\chi$:
\begin{eqnarray}\label{rhorec}
p_{M+1}&=&p_{M}|\langle 0|U(\tau)|0\rangle|^{2}+\langle 0|U(\tau)\chi_{M}U^{\dagger}(\tau)|0\rangle \\ \nonumber
\chi_{M+1}&=&p_{M}\Pi_{c}U(\tau)\Pi_{d}U^{\dagger}(\tau)\Pi_{c}+\Pi_{c}U(\tau)\chi_{M}U^{\dagger}(\tau)\Pi_{c}\:,
\end{eqnarray}
which can be iteratively solved. As the initial condition we choose
$p_{0}=1$ and the null matrix $\chi_{0} = 0$. % is given by the null matrix.
In passing we list the following properties of the map $\mathcal{M}$ without proof: $\mathcal{M}$ is a trace-preserving, completely positive and contractive linear map \cite{Hol}.
These properties imply that the spectrum of $\mathcal{M}$ is contained in the unit disk of the complex plane.

%%%%%%%%%%%%%%%%%%%%%%%%%%%%%%%%%%%%%%%%%%%%%%%%%%%%%%%%%%%%%%%%%%%%%%%%%%%%%%%%%%%%%%%%%%%%%%%
\begin{figure}[t]
\includegraphics[width=1\columnwidth]{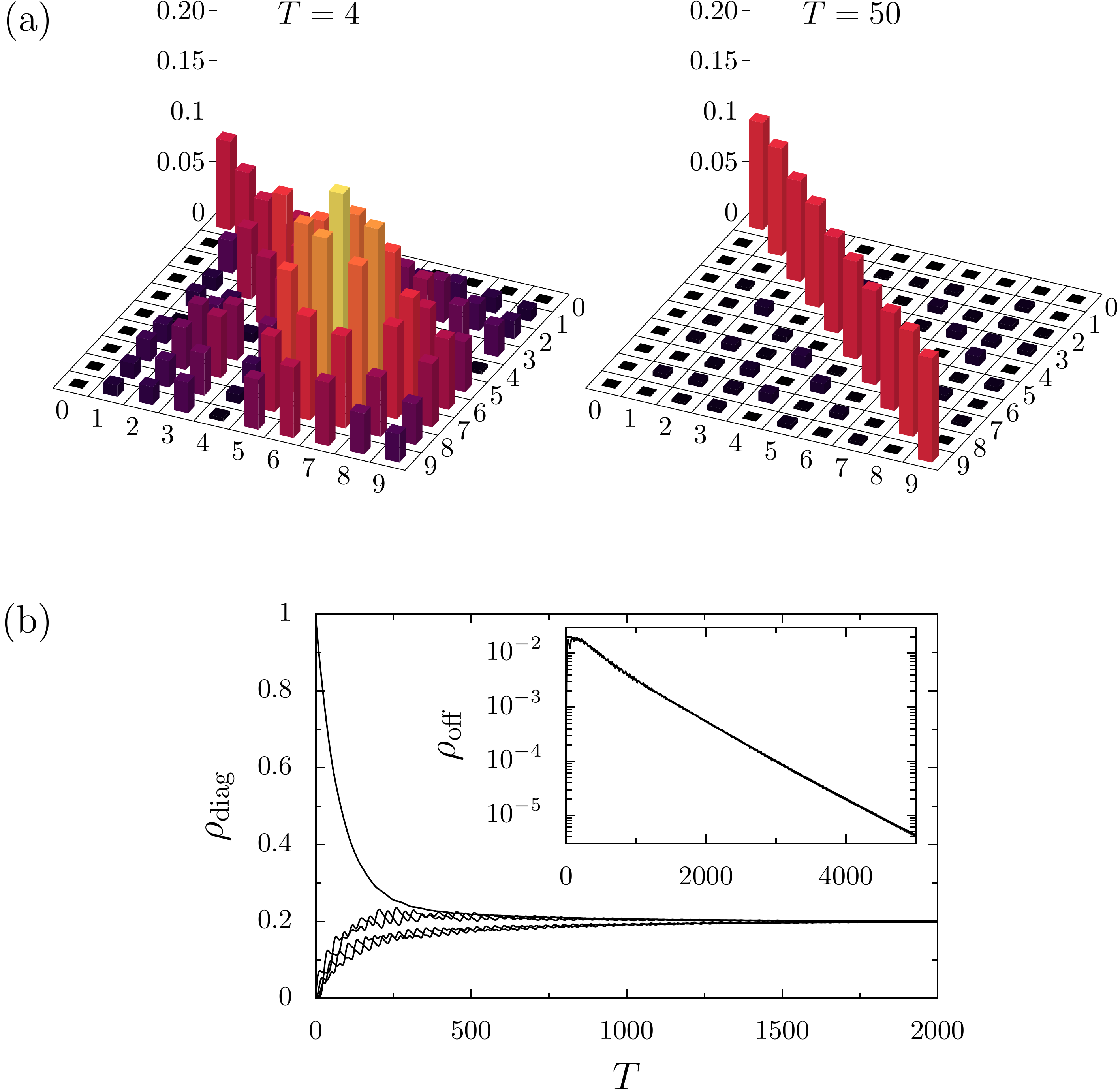}
 \caption{(Color online)~(a) The tomography of the density matrix
for a homogeneous chain~($\gamma= 1$, $\epsilon_{\ell}=0$) with $N=9$ sites,
displays the absolute values of the matrix elements at two times $T$.
The unitary time evolution of the system is interrupted by measurements at equal intervals of duration $\tau=1$.
In the left panel at the comparatively short time $T=4$ the diagonal elements are unevenly occupied and most of the non-diagonal elements have clearly non-zero values. The right panel illustrates the state reached at $T=50$. Then the density matrix has become almost diagonal with equal weights, $0.1$, and almost vanishingly small off-diagonal elements in good agreement with the fixed point solution~(\ref{fp}). (b) Density matrix  evolution for a dot coupled to a chain with $N=4$ sites, $\gamma=1$, and random on-site energies taken from a uniform distribution within $[-0.5,0.5]$. The diagonal components $\rho_{\text{diag}}$ converge to $0.2$, and
the average magnitude of the off-diagonal components, $\rho_{\text{off}}$ defined in the text, decays exponentially in time (see the inset).}\label{fig1}
\end{figure}
%%%%%%%%%%%%%%%%%%%%%%%%%%%%%%%%%%%%%%%%%%%%%%%%%%%%%%%%%%%%%%%%%%%%%%%%
%%%%%%%%%%%%%%%%%%%%%%%%%%%%%%%%%%%%%%%%%%%%%%%%%%%%%%%%%%%%%%%%%%%%%%%%%%%%%%%%%%%%%%%%%%%%%%%
\begin{figure}[t]
%\resizebox{7cm}{!}{
\includegraphics[width=1\columnwidth]{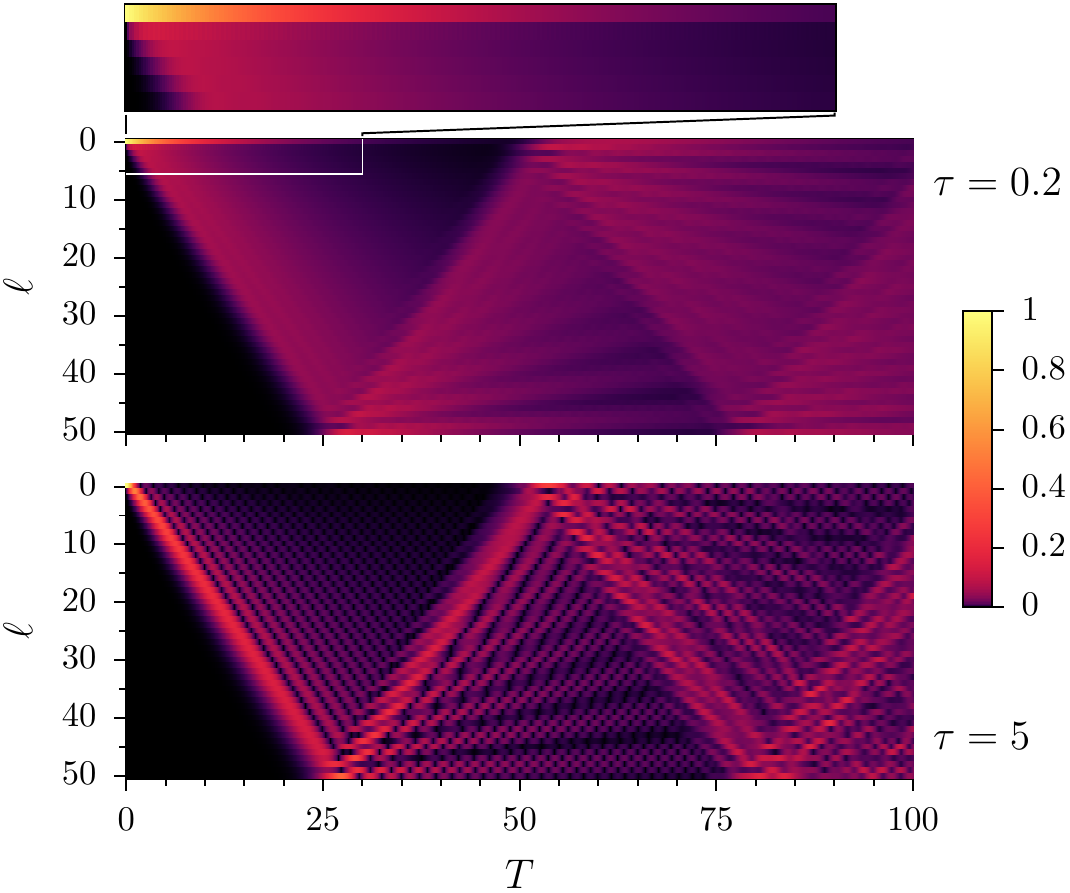}
%}
\caption{(Color online)~The diagonal elements of the density matrix in the position representation are encoded according to the scale given on the right as functions of position $\ell$ on the vertical axis and time $T$ on the horizontal axis. Here, a dot is coupled to a chain with 50 sites, $\epsilon_\ell=0$ and $\gamma =1$.  For
frequent measurements~($\tau =0.2$ in the upper panel), the Zeno effect is visible by the persistent population in the dot~(see also the enlargement), whereas in the lower panel for $\tau =5$ the dot population decays fast along the front ray that moves with the maximal group velocity 2 of the chain.
}
\label{fig2}
\end{figure}
%%%%%%%%%%%%%%%%%%%%%%%%%%%%%%%%%%%%%%%%%%%%%%%%%%%%%%%%%%%%%%%%%%%%%%%%

We present tomographic views of the density matrix $\rho_M$ in Fig.~\ref{fig1}(a) for different times $T=M \tau$ with the time between measurements $\tau =1$, as result of the $M$-fold application of the map $\mathcal{M}$. Here and in the sequel we measure time in units of $1/(\hbar \gamma_c)$.
After a few measurements, $T=4$, almost all elements of the density matrix are different from zero, including most of the non-diagonal ones. The diagonal elements assume different values indicating a non-uniform population of the dot and chain sites. At $T=50$, a stationary state with
almost vanishing non-diagonal and identical diagonal elements has been approached.
These results represent the typical behavior as also exemplified by the panel (b) of Fig.~\ref{fig1} visualizing the dynamics of a system with on-site energies
randomly chosen from a uniform distribution.
As a measure of the off-diagonal elements, $\rho_{i,j} =\langle i|\rho |j \rangle$, we determined the average of their absolute values
$\rho_{\text{off}}=\sum_{i\neq j}|\rho_{i,j}|/[N(N+1)]$, see the inset. Also in this case, we find the density matrix approaching the uniform stationary state, as in Fig.~\ref{fig1}(a) for identical on-site energies.

A more detailed account of the time-evolution of the diagonal elements is presented in Fig.~\ref{fig2} displaying the propagation of the population from the dot through the chain. In the upper panel with frequent measurements ($\tau =0.2$), the transfer of the population from the dot to the chain is hampered as a result of the Zeno effect, see also the enlarged detail on the top of the figure displaying the dot state and the first few chain sites. Behind a clearly visible front determined by the fastest signal propagation in the chain, a gradually decreasing excitation profile emerges. In contrast, for $\tau =5$ the population behind a pronounced  front ray decays fast on average with a superimposed periodic structure, which, at later times after the first reflections of the leading front, develops speckles. These speckles disperse only at much later times.
It is also notable that independent of the time $\tau$ the population dominantly propagates through the chain with the velocity $2$.
That can be understood from the fact that, apart from local effects at the border to the dot, the dynamics is dominated by the unitary time evolution in the chain.
For a chain decoupled from the dot, the
corresponding Hamiltonian (\ref{H}) with %letting
$\epsilon_{\ell}=0$ and $\gamma =0$, can be diagonalized to
yield the dispersion relation, ${\cal E}(k)=-2\cos k$, where $k=n\pi/N$ is the discrete momentum with $n=1,2,\cdots,N$. The group velocity is given by $v_{g}=\partial {\cal E}(k)/\partial k = 2\sin k$, and its maximum value is then $v_{g} = 2$. Hence it is the maximum group velocity that determines the front rays.

%%%%%%%%%%%%%%%%%%%%%%%%%%%%%%%%%%%%%%%%%%%%%%%%%%%%%%%%%%%%%%%%%%%%%%%%%%%%%%%%%%%%%%%%%%%%%%%
\begin{figure}[t]
%\resizebox{7cm}{!}{
\includegraphics[width=1\columnwidth]{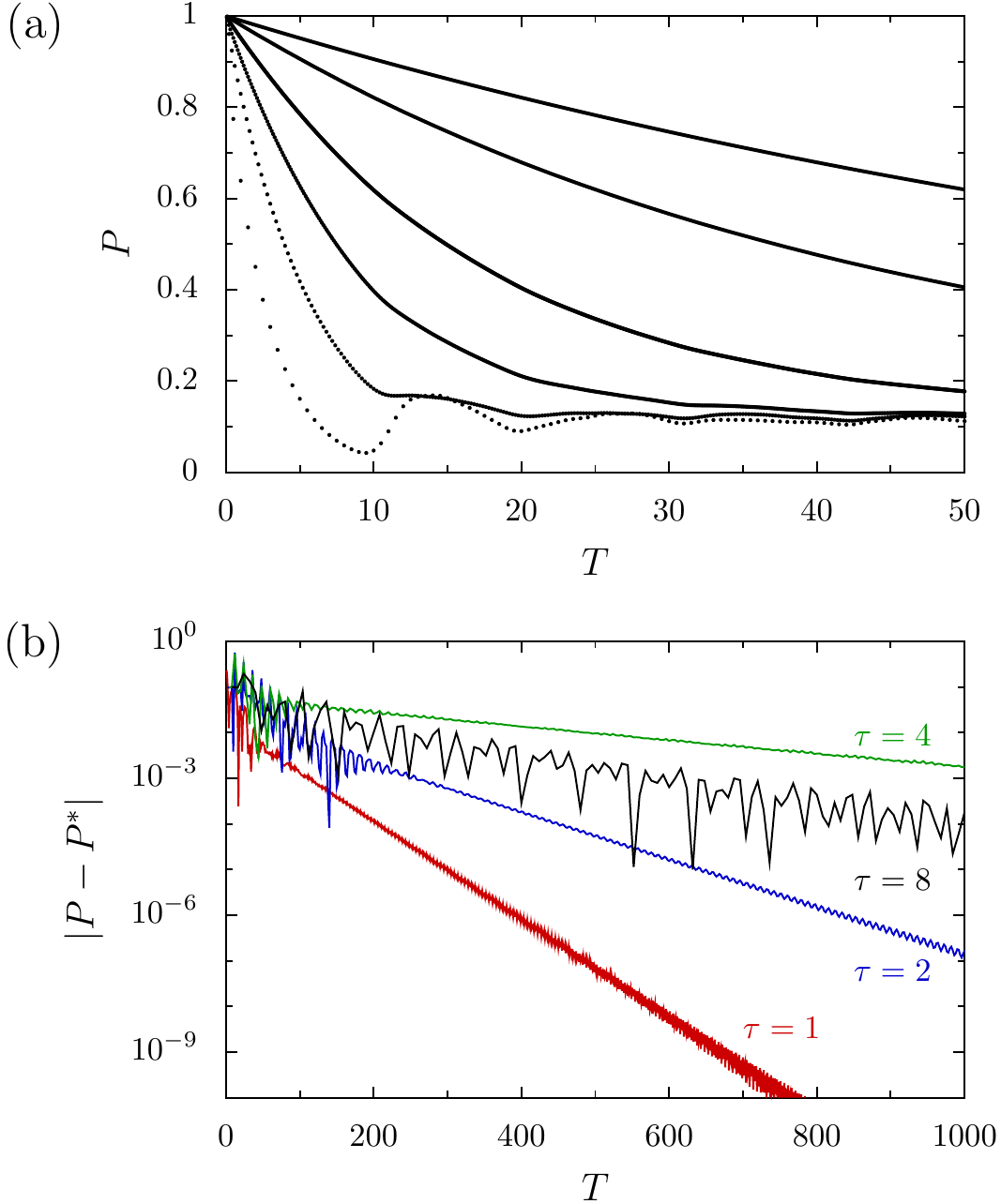}
%}
 \caption{
(Color online)~The decay of
the dot-state survival probability  $P = p_{T/\tau}$ in presence of a homogeneous chain with $N=9$ sites is displayed as a function of the time $T=M \tau$ after $M$ measurements for different values of $\tau$; the other parameters are the same as in Fig.~\ref{fig1}. Panel (a) presents frequent measurements with short intervals, $\tau=0.01,0.02,0.05,0.1,0.2,0.5$ from the upper to the lower curve. For these small $\tau$ values the decay becomes faster with increasing $\tau$. Panel (b) presents the difference of $P$ from the asymptotic state $P^* =0.1$ for larger values of $\tau$. In this regime the decay rate does no longer monotonically increase with $\tau$. In all cases at large times $T$ an average exponential decay sets in, which in some cases, is superimposed by an oscillatory behavior.
}
\label{fig3}
\end{figure}
%%%%%%%%%%%%%%%%%%%%%%%%%%%%%%%%%%%%%%%%%%%%%%%%%%%%%%%%%%%%%%%%%%%%%%%%

\section{Equal Stationary Probability}

The emergence of stationary states induced by repeated measurements was also reported for special systems~\cite{equilibration,similartoours1}, but up to now, neither conditions for their existence have been known, nor their universality has been noticed. In the following we address these questions.
The stationary solutions of the master equation (\ref{it}) are invariant under the action of the map $\mathcal{M}$ and therefore are solutions of 
$\mathcal{M} \rho_\infty = \rho_\infty$.  
One shows by inspection that
\begin{equation}\label{fp}
\rho_{\infty}=\frac{\openone}{N+1}\:,
\end{equation}
presents an invariant solution, where $\openone$ is the unit matrix.
In the Appendix~\ref{A} we prove that other solutions may only exist if there is an invariant chain state, i.e. a state with $\cal{M} \rho = \rho$ and $\Pi_d \rho = \rho \Pi_d = 0$.

This situation is exceptional; typically any chain state will evolve a dot component upon its unitary evolution. %(I WILL EXPAND THIS PROOF SOMEWHAT)
We note that the proof of the uniqueness of (\ref{fp}) holds true for general systems with finite-dimensional Hilbert space provided the measured observable possesses a non-degenerate eigenstate. The approach to the universal state (\ref{fp}) therefore is not restricted to the special model (\ref{H}).
Most notably, the uniform state (\ref{fp}) corresponds to the thermodynamic equilibrium at infinite
temperature. Apparently the repeated process of measurements ``heats
up'' the system to an extreme degree.

\section{Dot Population}

The relevant aspects of the dynamics in approaching this extreme state
are displayed by the time evolution of the survival probability of the dot population which is experimentally easier to access than the whole density matrix.
After $M$ measurements, this survival probability is given by the dot matrix element
$p_{M}=\langle 0|\rho_{M}|0\rangle$. According to (\ref{it}) and (\ref{M}), $p_M$ does not only include the uninterrupted sojourn on the dot, but also allows for those cases in which the particle leaves the dot and reenters it at a later time.

For a reservoir with a single level,
that is, $N=1$, analytic results are available. Since then $\Pi_{c}=|1\rangle \langle 1|$, the survival probability is simply given by the products of the $2\times 2$ transition matrix ${\mathbf T}$ as
\begin{equation}
p_{M}=(\mathbf{T}^{M})_{11}
\end{equation}
with %the transition matrix ${\mathbf T}$
\begin{equation}\label{T}
{\mathbf T}=\left( \begin{array}{cc}
                              T_{00} & T_{01} \\
                               T_{10} & T_{11}
                               \end{array}\right).
                               \end{equation}
where $T_{i,j}=|\langle i | U(\tau) |j\rangle |^{2}$. The unitary time evolution operator $ U(\tau)$
of this two-level system can be easily solved to yield $T_{00}=1-(\gamma/\hbar\Omega)^{2}\sin^{2}(\Omega \tau)$,
where the oscillation frequency is given by $\Omega =\hbar^{-1}\sqrt{\gamma^{2}+\epsilon^{2}/4}$. Using
$T_{01}=T_{10}$ resulting from the unitarity of $U(\tau)$, and the probability conservation of the unitary evolution, $T_{00}+T_{10}=T_{11}+T_{01}=1$,
upon diagonalizing ${\mathbf T}$ we obtain
\begin{equation}\label{ps1}
p_{M}=\frac{1}{2}[1+(2T_{00}-1)^{M}].
\end{equation}
For the measurement with a very short time interval, $\Omega \tau \ll 1$, 
one finds $T_{00}\approx 1 + {\cal O}(\tau^{2})$, and accordingly, $p_{M}\approx 1$ as long as $M$ is not too large.
This is a manifestation of the quantum Zeno effect caused by frequent measurements that obstruct the time evolution
of a quantum state. For larger values of $\tau$ or after a large number of measurements, $M\tau^{2}$ is no longer negligible; consequently $p_{M}$ then deviates from unity and
eventually converges to $1/2$, unless the measurement time interval accidentally exactly matches with the
period, $2\pi/\Omega$, for which $T_{00}$ is one, see the expression for $T_{00}$ below Eq. (\ref{T}).

For
reservoirs with more than one state numerical results are presented in Fig.~\ref{fig3}(a) where the survival probability
$P\equiv p_{T/\tau}$ is depicted as a function of the elapsed time $T=M \tau$ for relatively frequent measurements with accordingly small $\tau$. The smaller
the time interval between the measurements is, the slower the
decay of the survival probability becomes, in accordance with the
Zeno effect. For instance, for the uppermost curve ($\tau=0.01$), the survival
probability stays close to unity even after $10^{3}$ measurements.
On the other hand, for $\tau=0.5$, the decay becomes more rapid
and $P$ quite soon converges towards its asymptotic value $1/10$ which is in agreement with the survival probability resulting from (\ref{fp}) for $N=9$.

The panel (b) of Fig.~\ref{fig3} displays the deviation of the
survival probability from its asymptotic value $P^{*}=0.1$ in the
limit of large times. The decay rate sensitively depends on the
time interval $\tau$
although for larger values of $\tau$ it no longer grows monotonically with increasing $\tau$ values.
In general, the approach to the asymptotic value %was found to sensitively
crucially depends on the system details such as the size $N$ and the coupling strength $\gamma$.
%%%%%%%%%%%%%%%%%%%%%%%%%%%%%%%%%%%%%%%%%%%%%%%%%%%%%%%%%%%%%%%%%%%%%%%%%%%%%%%%%%%%%%%%%%%%%%%
\begin{figure}[t]
%\resizebox{7cm}{!}{
\includegraphics[width=1\columnwidth]{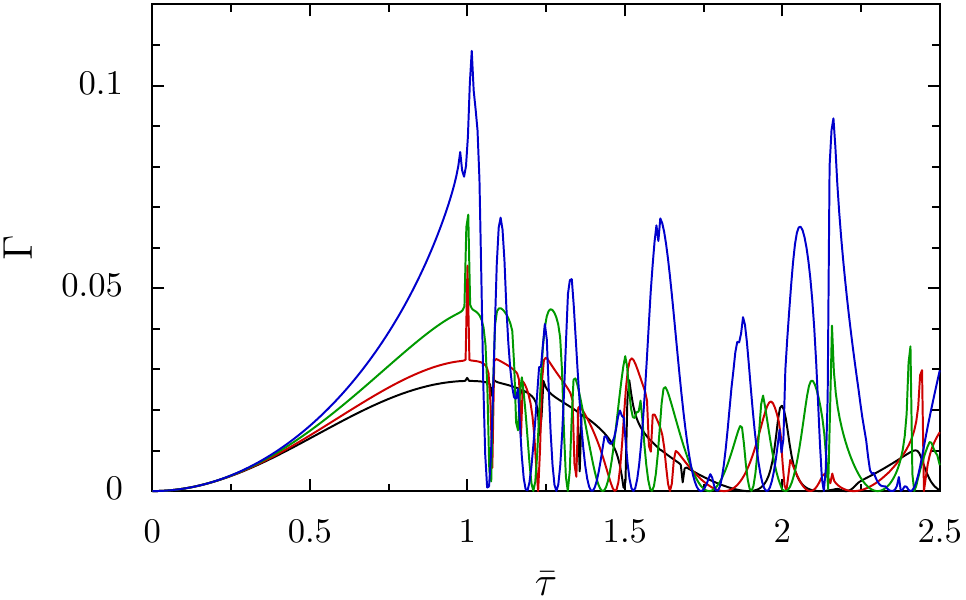}
%}
 \caption{
(Color online)~The rate $\Gamma = -\ln (|\lambda_1|)/\gamma^2$ determined by %related to
the eigenvalue $\lambda_1$ that in absolute value is closest to 1 is displayed as a function of the scaled time $\tilde{\tau} = \tau /\tau^*$ for different values of the coupling constant $\gamma= 1,\;0.75,\;0.5,\; 0.25$. Here $\tau^*$ denotes the shortest time scale of the unitary motion. All curves start out from zero displaying a universal $\tilde{\tau}^2$ growth behavior up to $\tilde{\tau}\approx 0.4$. Around  $\tilde{\tau}=0.75$ the $\Gamma$ values decrease with decreasing $\gamma$. Above  $\tilde{\tau}=1$ the rate $\Gamma$ presents an erratic behavior with large variations upon small parameter changes.
}
\label{fig4}
\end{figure}
%%%%%%%%%%%%%%%%%%%%%%%%%%%%%%%%%%%%%%%%%%%%%%%%%%%%%%%%%%%%%%%%%%%%%%%%

 \section{Decay Spectrum}\label{DS}

 The survival probability $p_M$ can be expressed in terms of the eigenvalues $\lambda_n$ and the  eigenfunctions $\Phi_n$ and $\varphi_n$ of the linear map $\mathcal{M}$ and the corresponding dual map $\mathcal{M}^+$, respectively, solving the eigenvalue problems $\mathcal{M} \Phi_n = \lambda_n \Phi_n$ and $\mathcal{M}^+ \varphi_n = \lambda_n \varphi_n$ . The dual map is defined such that $\text{Tr} \varphi \mathcal{M} \Phi = \text{Tr} \Phi \mathcal{M}^+ \varphi$ for all
matrices $\Phi$ and $\varphi$. Using the bi-orthogonality of the eigenfunction, $\text{Tr}\varphi_n \Phi_m = \delta_{n,m}$ and the initial condition $\rho_0=|0\rangle\langle 0|$ we obtain for the survival probability\cite{com}
\begin{equation}
\begin{split}
p_M &= \sum_n \lambda_n^M \langle 0|\varphi_n |0\rangle \langle 0|\Phi_n |0  \rangle\\
&= 1/(N+1) +\sum_{n\geq 1} \lambda_n^M \langle 0|\varphi_n |0 \rangle \langle 0|\Phi_n |0 \rangle \:,
\end{split}
\label{PL}
\end{equation}
where, in the second line, we split off the stationary part determined  by $\lambda_0=1$, $\varphi_0 = \openone$ and $\Phi_0=\openone/(N+1)$. The asymptotic decay towards the uniform stationary state is exponential with a decay rate given by the eigenvalue $\lambda_1$ whose absolute value is closest to unity. In practice though, the number of measurements to reach this asymptotic exponential decay regime may be extremely large due to the smallness of the coefficient $\langle 0|\varphi_1|0 \rangle \langle 0|\Phi_1|0 \rangle$ and the presence of other eigenvalues close to unity in absolute value. This may lead to extremely long transients with exponential decay at larger rates than those following from $\lambda_1$, or even non-exponential decay laws.

The eigenvalue $\lambda_1$ closest to unity depends in a regular way on the length of the interval $\tau$ as long as $\tau$ is shorter than the fastest time scale $\tau^* \approx 2\pi/\Delta E$ where $\Delta E$ is the width of the spectrum of the total Hamiltonian. The rate defined as the negative logarithm of $|\lambda_1|$, $\Gamma=-\ln (|\lambda_1|)/\gamma^2$, vanishes at $\tau =0$ and starts to grow proportionally to $\tau^2$, in accordance with the Zeno effect. As can be seen from Fig.~\ref{fig4} the scaling of the negative logarithm of $|\lambda_1|$ by $\gamma^2$ leads to a universal behavior of $\Gamma$ at small $\tau$ values. For values of $\tau$ larger than $\tau^*$ the dependence of the rate $\Gamma$ becomes very irregular and occasionally changes drastically from extremely small values corresponding to almost no decay of the survival probability to rather large values giving rise to rapid decay.

Finally we note without proof that a much simpler behavior emerges for an infinite chain with continuous energy spectrum and a spectral width which is much larger than any other energy scale. Under these conditions the dot population decays exponentially even in the absence of any measurement, i.e.
$|\langle 0|U(t)|0\rangle| = e^{-t/\tau_{0}}$ with $\tau_0 >0$. Then and only then a history in which the dot state is left at some time will never show a recurrence of the dot population, or mathematically expressed $\langle 0|U(\tau)\Pi_{c}U(\tau)|0\rangle =0$. Only in this case the traditional treatment of the Zeno effect yields the actual survival probability at a time $t$ taking into account the full histories up to this time.

\section{Summary and Remark}

We demonstrated that many measurements that are separated by finite periods of unitary time evolution lead to the universal infinite temperature state of equal probability. We expect this behavior also to apply for more general systems with finite dimensional Hilbert space if they are subject to a large number of measurements of the occupancy of part of the system.
This poses the question
whether this result is a mere artifact of von Neumann's instantaneous reduction postulate, on which our analysis was based, or whether the repeated interaction with a measuring device drives the system into this extreme state.

This work was supported by the National Research Grant funded by the Korean Government (NRF-2010-013-C00015).% and by the Deutsche Forschungsgemeinschaft %via the Grant No. HA 1517/28-1.
\appendix
\section{Proof of the  uniqueness of the infinite temperature state}
\label{A}
Any normalized solution of the stationary master equation
\begin{equation}
\rho = \cal{M} \rho
\label{rMr}
\end{equation}
with the master operator $\cal{M}$ given by Eq.~(\ref{M}) can be expressed as
\begin{equation}
\rho = \openone /(N+1) +x\:,
\label{1x}
\end{equation}
where $\text{Tr} x = 0$ because the first term of the right hand side of Eq.~(\ref{1x}) already takes into account the full normalization of the density matrix $\rho$.
We demonstrate that a nontrivial solution with $x \neq 0$ and hence, other than infinite temperature solutions, may only exist if the unitary time evolution $U(\tau)$ has invariant chain-states i.e. if
\begin{equation}
y = \Pi_c U(\tau) y U^\dagger(\tau) \Pi_c
\label{Co}
\end{equation}
has non-trivial solutions.
Under typical conditions this will not be the case. The formal proof of Eq. (\ref{Co}) as a necessary and sufficient condition is next presented.
Since $\openone/(N+1)$ solves Eq.~(\ref{rMr}) one obtains
\begin{equation}
x = {\cal M} x \:.
\label{xMx}
\end{equation}
Hence, in analogy to Eq.~(\ref{rhorec}) $x$ can be decomposed into a dot and a chain part reading
\begin{equation}
x = q \Pi_d + z\:,
\label{xdz}
\end{equation}
where $q$ is a real number and $z$ an operator on the chain sub-space, hence, satisfying $\Pi_d z = z \Pi_d =0$, with $\text{Tr} z = -q$. Using Eq.~(\ref{xMx}) one obtains coupled equations for the unknowns $d$ and $z$
\begin{align}
\label{z}
z&= \Pi_c U(\tau) z U^\dagger(\tau)\Pi_c + q \Pi_c U(\tau) \Pi_d U^\dagger(\tau) \Pi_c\\
\label{d}
q \Pi_d & = \Pi_d U(\tau) z U^\dagger(\tau) \Pi_d + q \Pi_d U(\tau) \Pi_d U^\dagger(\tau) \Pi_d \:.
\end{align}
We separately consider the two possible cases with $q=0$ and $q\neq 0$.

(a) For $q=0$ Eqs.~(\ref{z}) and (\ref{d}) simplify to read
\begin{align}
\label{z0}
z &=\Pi_c U(\tau) z U^\dagger(\tau) \Pi_c\\
\label{d0}
0 &=\Pi_d U(\tau) z U^\dagger(\tau) \Pi_d \:.
\end{align}
The first condition (\ref{z0}) has the form of Eq.~(\ref{Co}) and hence has solutions only under exceptional conditions on the unitary chain dynamics and the time $\tau$.
We note that Eq.~(\ref{d0}) is a consequence of Eq.~(\ref{z0}). This follows by taking the trace over both sides of Eq.~(\ref{z0}) and by replacing $z$ by $U(\tau) z U^\dagger(\tau)$ under the trace on the left hand side, yielding
\begin{equation}
\text{Tr}\left [ U(\tau) z U^\dagger(\tau) \right ] =   \text{Tr} \left [ \Pi_c U(\tau) z U^\dagger(\tau) \Pi_c \right ]
\end{equation}
and consequently
\begin{equation}
\text{Tr} \left [ \Pi_d  U(\tau) z U^\dagger(\tau) \right ] = 0
\label{eqC}
\end{equation}
which is equivalent to Eq.~(\ref{d0}) because $\Pi_d$ is the projection onto a one-dimensional subspace.

(b) If $q \neq 0$ one can divide both Eqs.~(\ref{z},\ref{d}) by $q$ and obtains
for $\bar{z} = z/q$ two inhomogeneous equations reading
\begin{align}
\label{zd}
\bar{z} &= \Pi_c U(\tau) \bar{z} U^\dagger(\tau) \Pi_c +\Pi_c U(\tau) \Pi_d U^\dagger(\tau) \Pi_c \\
\label{dd}
\Pi_d & = \Pi_d U(\tau) \bar{z} U^\dagger(\tau) \Pi_d + \Pi_d U(\tau) \Pi_d U^\dagger(\tau) \Pi_d \:.
\end{align}
By similar arguments as above one finds that the second equation follows from the first one. It is therefore sufficient to only consider the Eq.~(\ref{zd}).
One finds by inspection as solution of Eq.~(\ref{zd}) $\bar{z} = \Pi_c$. This however does not present an admissible solution because it does not satisfy the normalization condition $\text{Tr} z = -q$  implying $\text{Tr} \bar{z} =-1$ in contrast to
$\text{Tr}\bar{z} =\text{Tr} \Pi_c =N$. Whether there exist further solutions of the inhomogeneous equation (\ref{zd}) can be decided by means of the Fredholm alternative. It says that more than one solution of an inhomogeneous equation may only exist if the homogeneous part of the equation possesses nontrivial solutions and if the inhomogeneity is orthogonal to all solutions of the dual homogeneous problem, see Sec.~\ref{DS}. One can show that in the present case the condition for the inhomogeneity is automatically satisfied. The homogeneous part of Eq.~(\ref{zd}) again leads to a condition of the form of Eq.~(\ref{Co}). Assuming the existence of a nontrivial normalized solution $u$ ($\text{Tr} u =1$) of (\ref{Co}) we obtain for the general stationary solution of Eq.~(\ref{rMr}) the expression
\begin{equation}
\rho = \left ( \frac{1}{1+N} +q \right ) \openone - q (N+1)u\:,
\end{equation}
where $u$ is a chain-state operator ($\Pi_d u = u \Pi_d =0$) and $q$ can assume arbitrary values in $(-1/(N+1),N)$.

Hence in both cases (a) and (b) the uniqueness of the solution of the stationary master equation is guaranteed if and only if Eq. (\ref{Co})
only has the trivial solution $y=0$.
.


\begin{thebibliography}{99}
\bibitem{com} For the sake of simplicity we assume that $\mathcal{M}$ has a spectral representation.
\end{thebibliography}
\end{document}